\documentclass[pra,showpacs,showkeys,amsfonts,amsmath,twocolumn]{revtex4}
\usepackage{bm}
\usepackage{graphicx}
\RequirePackage{mathptm}

\numberwithin{equation}{section}
\newcommand{\si}[1]{\sigma_{#1}}
\newcommand{\sa}[2]{\sigma_{#1}^{#2}}

\newcommand{\W}[4]{\begin{cases}
#1 ,&#2\\
#3 ,&#4
\end{cases}}
\newcommand{\ro}{\rho}

\newcommand{\ga}[1]{\gamma_{#1}}

\newcommand{\I}{\openone}
\newcommand{\ket}[1]{|{#1}\rangle}
\newcommand{\bra}[1]{\langle {#1} |}

\newcommand{\C}{\mathbb C}

\newcommand{\mr}[1]{\mathrm{#1}}
\newcommand{\e}[1]{{\boldsymbol e}_{#1}}

\begin{document}
\title{Quantum interference and evolution of entanglement in a system of three-level atoms}
\author{{\L}ukasz Derkacz}
\affiliation{Institute of Theoretical Physics\\ University of
Wroc{\l}aw\\
Pl. M. Borna 9, 50-204 Wroc{\l}aw, Poland}
\author{Lech Jak{\'o}bczyk\footnote{
E-mail address: ljak@ift.uni.wroc.pl}} \affiliation{Institute of
Theoretical Physics\\ University of
Wroc{\l}aw\\
Pl. M. Borna 9, 50-204 Wroc{\l}aw, Poland}
\begin{abstract}
We consider a pair of three-level atoms interacting with the vacuum.
The process of disentanglement due to spontaneous emission and the
role of quantum interference between principal transitions in this
process, are analysed. We show that the presence of interference can
slow down disentanglement. In the limit of maximal interference,
some part of initial entanglement can survive.
\end{abstract}
\maketitle
\section{Introduction}
The process of decoherence and degradation of entanglement in open
quantum systems was studied mainly in the case of  pair of two-level
atoms interacting with vacuum or thermal noises. This system is
simple to study, since the amount of its entanglement can be
quantified by  concurrence \cite{wootters}. This quantity
discriminates between separable and entangled states and is
analytically computable. The dynamics of the system given in the
Markovian regime by the master equation, can be used to obtain time
evolution of concurrence and to analyse the influence of the process
of decoherence on entanglement. Let us mention two particular
results of such studies. The process of disentanglement was
considered in the case of a pair of atoms, each interacting with its
own reservoir at zero temperature. It turns out that concurrence of
initially entangled state can vanish asymptotically or in finite
time, depending on initial state \cite{eberly, dodd, jam}. This
behaviour should be compared with asymptotic decoherence of any
initial state. On the other hand, individual atoms located inside
two independent environments at infinite temperatures, always
disentangle at finite times \cite{jajam}.
\par
In the present paper, we extend this kind of studies to the case of
pair of three-level atoms. The analysis is much more involved since
there is no simple necessary and sufficient criterion of
entanglement for a pair of $d$-level systems with $d\geq 3$. Peres'
separability criterion \cite{peres} only shows that states which are
not positive after partial transposition (NPPT states) are
entangled. But there can exist entangled states which are positive
after this operation \cite{horodecki} (PPT states). Using this
criterion we at most  can study the evolution of NPPT states and ask
when they become PPT states, but in the models considered in the
present paper we can directly observe when stationary asymptotic
states are separable.
\par
Interesting example of dissipative dynamics comes from the analysis
of three-level V-type atomic system where spontaneous emission may
take place from two excited levels to the ground state and direct
transition between excited levels is not allowed. However the
indirect coupling between excited states can appear due to
interaction with the vacuum (quantum interference) \cite{agarwal}.
This interference results from the following mechanism: spontaneous
emission from one transition modifies spontaneous emission from the
other transition \cite{ficek}. There were many studies on the effect
of quantum interference on various processes including: resonance
fluorescence \cite{heg}, quantum jumps \cite{zoller}, the presence
of ultranarrow spectral lines \cite{swain} or amplification without
population inversion \cite{harris}.
\par
In our research we consider a pair of such three-level systems and
study another aspect of quantum interference, namely its influence
on the evolution of entangled states. In the case of distant atoms
we expect that dissipation causes disentanglement, but the rate of
this process may depend on interference. It is indeed true, as shows
analysis of the models considered in the paper. We demonstrate it
for some classes of pure and mixed states by proving  that the
measure of entanglement vanishes asymptotically, but the larger is
the effect of interference, the slower is the process of
disentanglement. In the limit of maximal interference we obtain very
interesting phenomenon: some part of initial entanglement can
survive and we obtain  asymptotic states with non-zero entanglement
despite of of the inevitable process of decoherence.
\par
In this paper we study two models of three-level systems with
quantum interference between transitions. The simpler, proposed in
Ref. \cite{ficek}, applied to the pair of atoms can be solved
analytically. This model simulates many aspects of V-type
three-level atom where interference effect was discovered
\cite{agarwal}. We study the dynamics of a pair of such atoms
numerically and show that both models predict similar behaviour of
entanglement.

\section{Three-level atom and quantum interference}
Consider a three-level atom in the V configuration. The atom has two
nondegenerate excited states $\ket{1}$, $\ket{2}$ and the ground
state $\ket{3}$. Assume that excited levels can decay to the ground
state by spontaneous emission, whereas a direct transition between
excited levels is not allowed. If the dipole moments of these two
transitions are parallel, then indirect coupling between states
$\ket{1}$ and $\ket{2}$ can appear due to interaction with the
vacuum (quantum interference between transitions $\ket{1}\to
\ket{3}$ and $\ket{2}\to \ket{3}$) (FIG. 1).
\par
\begin{figure}[h]
\centering
{\includegraphics[height=55mm]{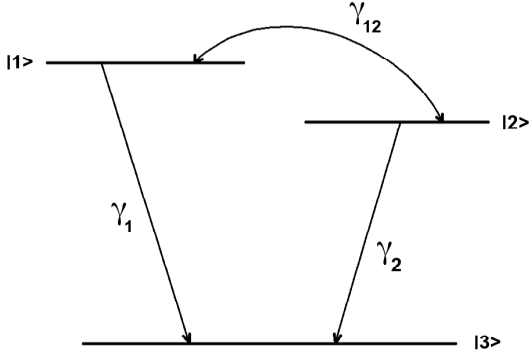}}\caption{Three-level
atom in the $V$ configuration (system I)}
\end{figure}
Time evolution of such system (which we call system I) is given by
the master equation \cite{ficek}
\begin{equation}
\frac{d\rho}{dt}=-i[H,\rho]+L_{I}\rho\label{me}
\end{equation}
where the damping term is
\begin{equation}
\begin{split}
L_{I}\rho&=\frac{1}{2}\,\ga{1}(2\si{31}\rho\si{13}-\si{11}\rho-\rho\si{11})\\
&+\frac{1}{2}\,\ga{2}(2\si{32}\rho\si{23}-\si{22}\rho-\rho\si{22})\\
&+\frac{1}{2}\,\ga{12}(2\si{31}\rho\si{23}-\si{21}\rho-\rho\si{21})\\
&+
\frac{1}{2}\,\ga{12}(2\si{32}\ro\si{13}-\si{12}\rho-\rho\si{12})\label{gen1}
\end{split}
\end{equation}
In this equation, $\si{jk}$ is the transition operator from
$\ket{k}$ to $\ket{j}$ and $\ga{1},\,\ga{2}$ are spontaneous
emission constants of $\ket{1}$ and $\ket{2}$ to the ground level
$\ket{3}$. In addition
\begin{equation}
\ga{12}=\beta_{I} \sqrt{\ga{1}\ga{2}}\label{g12}
\end{equation}
gives cross damping term between $\ket{1}\to \ket{3}$ and
$\ket{2}\to \ket{3}$. The parameter $\beta_{I}$ represents the
mutual orientation of transition dipole moments: when $\beta_{I}=1$,
quantum interference is maximal and for $\beta_{I}=0$ it vanishes.
Since we are mainly interested in the evolution of initial states
due to the spontaneous emission, the form of the Hamiltonian $H$ is
not discussed (see \cite{ficek} for details).
\par
\begin{figure}[h]
\centering
{\includegraphics[height=55mm]{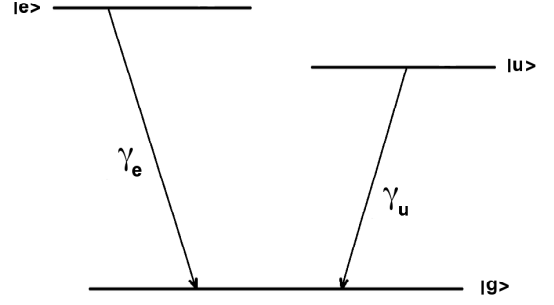}}\caption{Three-level
atom in the $V$ configuration (system II)}
\end{figure}
It is known that in atoms used in atomic spectroscopy, the
transition dipole moments are usually perpendicular, so the system
(I) described by equation (\ref{gen1}) is difficult to realize.
However, as was shown in Ref. \cite{ficek}, the effects of quantum
interference can be duplicated to a large degree by the three-level
system (II) (FIG. 2) with  exited states $\ket{e},\, \ket{u}$ and
the ground state $\ket{g}$, evolving according to the master
equation with damping operator
\begin{equation}
\begin{split}
L_{II}\rho&=\frac{1}{2}\ga{e}\,(2\si{ge}\rho\si{eg}-\si{ee}\rho-\rho\si{ee})\\
&+\frac{1}{2}\ga{u}\,(2\si{gu}\rho\si{ug}-\si{uu}\rho-\rho\si{uu})\label{gen2}
\end{split}
\end{equation}
In the system (II), the dipole moments can be perpendicular and the
measure of quantum interference is given by
\begin{equation}
\beta_{II}=\frac{\ga{e}-\ga{u}}{\ga{e}+\ga{u}} \label{beta}
\end{equation}
So we can expect maximal effects of quantum interference in the
system (II) when the level $\ket{u}$ is metastable.

\section{Entangled pair of three-level atoms}
Now we consider a pair of three-level systems in the V
configuration. In the context of evolution given by master equation
generated by (\ref{gen1}) or (\ref{gen2}), we address the following
question: how the dynamics of compound system of two atoms prepared
initially  in the entangled  state is influenced by the presence of
quantum interference? In particular, we  may look at the time
evolution of the appropriate measure of entanglement. Since damping
causes that pure states evolve into mixed states, we need effective
measure of mixed - state entanglement. As such measure one usually
takes entanglement of formation $E_{F}(\rho)$ \cite{bennet}, but in
practice it is not known how to compute this measure for mixed
states of pairs of $d$ - level systems in the case when $d>2$. A
computable measure of entanglement was proposed in Ref.
\cite{werner}. It is based on the trace norm of the partial
transposition $\rho^{T_{A}}$ of the state $\rho$. From the Peres'
criterion of separability \cite{peres}, it follows that if
$\rho^{T_{A}}$ is not positive, then $\rho$ is not separable. So one
defines \textit{negativity} of the state $\rho$ as
\begin{equation}
N(\rho)=\frac{||\rho^{T_{A}}||_{1}-1}{2}\label{neg}
\end{equation}
$N(\rho)$ equals to the absolute value of the sum of negative
eigenvalues of $\rho^{T_{A}}$ and  is an entanglement monotone
\cite{werner}, but it can not detect bound entangled states
\cite{horodecki}.
\par
The compound system of two atoms is defined  on the space
$\C^{3}\otimes \C^{3}$ and the matrix elements of the state $\rho$
will be considered with respect to the basis
\begin{equation}
\begin{split}
\{&\e{1}\otimes\e{1},\,
\e{1}\otimes\e{2},\,\e{1}\otimes\e{3},\,\e{2}\otimes\e{1},\,\e{2}\otimes\e{2},\,
\e{2}\otimes\e{3},\\
&\e{3}\otimes\e{1},\,\e{3}\otimes\e{2},\,\e{3}\otimes\e{3}\}\label{basis}
\end{split}
\end{equation}
where $\{\e{1},\, \e{2},\, \e{3}\}$ is the canonical basis of
$\C^{3}$. In our discussion of entanglement evolution we focus on
two classes of initial states. The first one consist of pure states
of the form
\begin{equation}
\Psi_{\theta,
\varphi}=\cos\theta\sin\varphi\,\e{1}\otimes\e{1}+\sin\theta\sin\varphi\,\e{2}\otimes\e{2}+\cos\varphi\,
\e{3}\otimes\e{3}\label{purestates}
\end{equation}
where $\theta,\,\varphi \in [0,\pi/2]$. The states
(\ref{purestates}) have negativity
\begin{equation}
N(\Psi_{\theta,\varphi})=\cos\theta\sin\varphi\cos\varphi
+\cos\theta\sin\theta\sin^{2}\varphi+\sin\theta\cos\varphi\sin\varphi\label{negpure}
\end{equation}
In the special case $\theta=\pi/4,\, \varphi=\arccos
\frac{1}{\sqrt{3}}$, we obtain the state
\begin{equation}
\Psi_{\mr{max}}=\frac{1}{\sqrt{3}}\,(\e{1}\otimes\e{1}+\e{2}
\otimes\e{2}+\e{3}\otimes\e{3})\label{max}
\end{equation}
with maximal negativity $N(\Psi_{\mr{max}})=1$. The second class of
so called \textit{isotropic} mixed states \cite{hhh}, is defined as
follows: let
$$
\rho_{0}=\frac{\I_{9}}{9}
$$
where $\I_{9}$ is the identity matrix in $\C^{9}$, be the maximally
mixed state of two three - level systems, then
\begin{equation}
W=(1-p)\,\rho_{0}+p\,\ket{\Psi_{\mr{max}}}\bra{\Psi_{\mr{max}}},\quad
p\in [0,1]\label{W}
\end{equation}
interpolate between maximally entangled and maximally mixed states.
Notice that states (\ref{W}) are the natural generalizations of
Werner states \cite{wer} of two-qubit systems. One can check that
negativity of (\ref{W}) equals
\begin{equation}
N(W)=\W{0}{p\leq
\frac{1}{4}}{\frac{1}{3}\,(4p-1)}{p>\frac{1}{4}}\label{negW}
\end{equation}
\section{Time evolution of negativity}
Consider first the system of two distant three - level atoms $A$ and
$B$ , both of type (II). This case is simpler to analyse and can be
solved exactly. Since the atoms are independent and do not interact,
the dissipative part of dynamics is given by the master equation
\begin{equation}
\frac{d\rho}{dt}=L_{\mr{II}}^{AB}\rho\label{meII}
\end{equation}
with
\begin{equation}
\begin{split}
L_{II}^{AB}\rho&=\frac{1}{2}\ga{e}\,(2\sa{ge}{A}\rho\sa{eg}{A}-
\sa{ee}{A}\rho-\rho\sa{ee}{A})\\
&+\frac{1}{2}\ga{u}\,(2\sa{gu}{A}\rho\sa{ug}{A}-\sa{uu}{A}\rho-\rho\sa{uu}{A})\\
&+\frac{1}{2}\ga{e}\,(2\sa{ge}{B}\rho\sa{eg}{B}-\sa{ee}{B}\rho-\rho\sa{ee}{B})\\
&+\frac{1}{2}\ga{u}\,(2\sa{gu}{B}\rho\sa{ug}{B}-\sa{uu}{B}\rho-\rho\sa{uu}{B})
\end{split}\label{gen2AB}
\end{equation}
where
$$
\sa{kl}{A}=\si{kl}\otimes \I_{3},\quad \sa{kl}{B}=\I_{3}\otimes
\si{kl},\quad k,l=e,u,g
$$
Moreover, we identify the states $\ket{e},\, \ket{u}$ and $\ket{g}$
with vectors $\e{1},\, \e{2}$ and $\e{3}$ respectively. The master
equation (\ref{meII}) can be used to obtain the equations for matrix
elements  of any density matrix of compound system, with respect to
the basis (\ref{basis}). The calculations are tedious but
elementary, so we are not discussing details. As a result, we obtain
the following properties of dynamics given by (\ref{meII}):
\begin{enumerate}
\item for $\ga{u}>0$, the dynamics brings all initial states into
unique asymptotic state $\ket{g}\otimes \ket{g}$,
\item in the case of maximal interference ($\ga{u}=0$), the dynamics
is not ergodic, and asymptotic stationary states $\rho_{\mr{as}}$
depend on initial conditions.
\end{enumerate}
The case of maximal interference will be studied in the next
section, now we discuss the evolution of entanglement in the first
case. Since the asymptotic state is separable, negativity of any
initially entangled state goes to zero with time. To see some
details, consider initial states of the form
\begin{equation}
\rho=
\begin{pmatrix}
\rho_{11}&\cdots&\rho_{15}&\cdots&\rho_{19}\\
\vdots&&\vdots&&\vdots\\
\rho_{51}&\cdots&\rho_{55}&\cdots&\rho_{59}\\
\vdots&&\vdots&&\vdots\\
 \rho_{91}&\cdots&\rho_{95}&\cdots&\rho_{99}
\end{pmatrix}\label{initial}
\end{equation}
Notice that pure states (\ref{purestates}) are particular examples
of (\ref{initial}). The solution $\rho(t)$ of master equation
(\ref{meII}) for initial condition (\ref{initial}) has matrix
elements
\begin{equation}
\begin{split}
\rho_{11}(t)&=e^{-2\ga{e}t}\,\rho_{11}\\
\rho_{15}(t)&=e^{-(\ga{e}+\ga{u})t}\,\rho_{15}\\
\rho_{19}(t)&=e^{-\ga{e}t}\,\rho_{19}\\
\rho_{33}(t)&=(e^{-\ga{e}t}-e^{-2\ga{e}t})\,\rho_{11}\\
\rho_{55}(t)&=e^{-2\ga{e}t}\,\rho_{55}\\
\rho_{59}(t)&=e^{-\ga{u}t}\,\rho_{59}\\
\rho_{66}(t)&=(e^{-\ga{u}t}-e^{-2\ga{u}t})\,\rho_{55}\\
\rho_{77}(t)&=(e^{-\ga{e}t}-e^{-2\ga{e}t})\,\rho_{11}\\
\rho_{88}(t)&=(e^{-\ga{u}t}-e^{-2\ga{u}t})\,\rho_{55}\\
\rho_{99}(t)&=(1+e^{-2\ga{e}t}-2e^{-\ga{e}t})\,\rho_{11}\\
&+(1+e^{-2\ga{u}t}-2e^{-\ga{u}t})\,\rho_{55}+\rho_{99}
\end{split}
\end{equation}
where for remaining matrix elements one can use hermiticity of
$\rho(t)$. Even in that case, the analytic formula for negativity as
a function of time is rather involved, so we consider explicit
example. Take as initial state $\Psi_{\mr{max}}$ which is maximally
entangled, then
\begin{equation}
N_{\Psi_{\mr{max}}}(t)=\frac{1}{3}\left(e^{-2\ga{e}t}+e^{-(\ga{e}
+\ga{u})t}+e^{-2\ga{u}t}\right)
\label{negpsit}
\end{equation}
\begin{figure}[t]
\centering
{\includegraphics[height=68mm,angle=270]{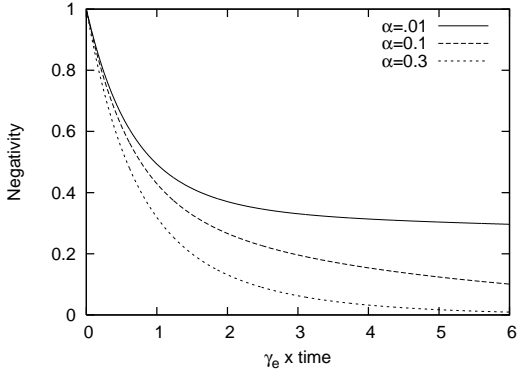}}\caption{Time-dependence
of negativity for initial state $\Psi_{\mr{max}}$ for different
values of $\alpha=\ga{u}/\ga{e}$ }
\end{figure}
Observe that (\ref{negpsit}) vanishes asymptotically at the rate
depending on degree of interference: the larger is the effect of
interference, the slower is the process of disentanglement (FIG. 3).
Notice also that in the limit of maximal interference ($\ga{u}\to
0$), some part of initial entanglement can survive. One can also
check that other entangled states from the class (\ref{purestates})
behave similarly (see FIG. 4).
\begin{figure}[t]
\centering
{\includegraphics[height=68mm,angle=270]{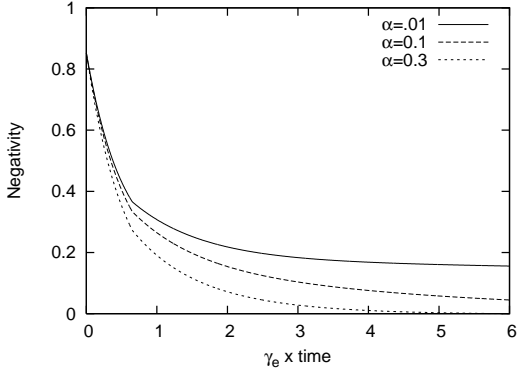}}\caption{Time-dependence
of negativity for initial state (\ref{purestates}) with
$\theta=\pi/8,\, \varphi=\pi/6$ and different values of
$\alpha=\ga{u}/\ga{e}$ }
\end{figure}
\begin{figure}[b]
\centering
{\includegraphics[height=68mm,angle=270]{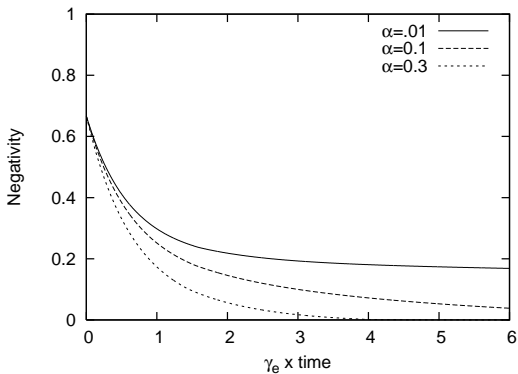}}\caption{Time-dependence
of negativity for initial state $W$ with $p=3/4$ and different
values of $\alpha=\ga{u}/\ga{e}$ }
\end{figure}
When the initial state is mixed isotropic state $W$ given by
(\ref{W}), then $W(t)$ has matrix elements
\begin{equation}
\begin{split}
W_{11}(t)&=\frac{1}{9}(1+2p)\,e^{-2\ga{e}t}\\
W_{15}(t)&=\frac{1}{3}p\,e^{-(\ga{e}+\ga{u})t}\\
W_{19}(t)&=\frac{1}{3}p\,e^{-\ga{e}t}\\
W_{22}(t)&=\frac{1}{9}(1-p)\,e^{-(\ga{e}+\ga{u})t}\\
W_{33}(t)&=\frac{1}{9}(p-1)\,e^{-(\ga{e}+\ga{u})t}-
\frac{1}{9}(1+2p)\,e^{-2\ga{e}t}+\frac{1}{3}\,e^{-\ga{e}t}\\
W_{44}(t)&=\frac{1}{9}(1-p)\, e^{-(\ga{e}+\ga{u})t}\\
W_{55}(t)&=\frac{1}{9}(1+2p)\, e^{-2\ga{u}t}\\
W_{59}(t)&=\frac{1}{3}p\, e^{-\ga{u}t}\\
W_{66}(t)&=\frac{1}{9}(p-1)\,e^{-(\ga{e}+\ga{u})t}-
\frac{1}{9}(1+2p)\,e^{-2\ga{u}t}+\frac{1}{3}\,e^{-\ga{u}t}\\
W_{77}(t)&=W_{33}(t)\\
W_{88}(t)&=W_{66}(t)\\
W_{99}(t)&=1+\frac{1}{9}(1+2p)\,
(e^{-2\ga{e}t}+e^{-2\ga{u}t})+\frac{2}{9}(1-p)\,e^{-(\ga{e}+\ga{u})t}\\
&-\frac{2}{3}\,(e^{-\ga{e}t}+e^{-\ga{u}t})
\end{split}
\end{equation}
Explicit formula for negativity of $W(t)$ is lengthy and involved,
so we not reproduce it here. But it follows from the calculations,
that the time dependence of negativity is  similar to the case of
pure states (FIG. 5).
\section{Maximal interference and asymptotic entanglement}
The case of maximal interference  in system (II) can be discussed in
details. Take any initial state $\rho$ with matrix elements
$\rho_{ij}$. The dynamics given by the master equation (\ref{meII})
has a remarkable property: in the case of maximal interference
between transitions $\ket{e}\to \ket{g}$ and $\ket{u}\to \ket{g}$,
there are nontrivial stationary asymptotic states $\rho_{\mr{as}}$.
By a direct calculations, one can show that $\rho_{\mr{as}}$ have
the following non-vanishing matrix elements
\begin{equation}
\begin{split}
&(\rho_{\mr{as}})_{55}=\rho_{55},\quad
(\rho_{\mr{as}})_{56}=\rho_{56},\quad
(\rho_{\mr{as}})_{58}=\rho_{58},\quad\\
&(\rho_{\mr{as}})_{59}=\rho_{59},\quad
(\rho_{\mr{as}})_{68}=\rho_{68},\quad
(\rho_{\mr{as}})_{66}=\rho_{44}+\rho_{66}\\
&(\rho_{\mr{as}})_{69}=\rho_{47}+\rho_{69},\quad
(\rho_{\mr{as}})_{88}=\rho_{22}+\rho_{88}\\
&(\rho_{\mr{as}})_{89}=\rho_{23}+\rho_{89},\quad
(\rho_{\mr{as}})_{99}=\rho_{11}+\rho_{33}+\rho_{77}+\rho_{99}
\end{split}
\end{equation}
It is very interesting that although two atoms are independent and
do not interact, some of these asymptotic states can be entangled.
Quantum interference causes that part of initial entanglement can
survive the process of decoherence. For example, if initial states
are pure states (\ref{purestates}), then $\rho_{\mr{as}}$ has only
four nonvanishing matrix elements
\begin{equation}
\begin{split}
&(\rho_{\mr{as}})_{55}=\sin^{2}\varphi\sin^{2}\theta,\quad
(\rho_{\mr{as}})_{59}=\cos\varphi\sin\varphi\sin\theta,\\
&(\rho_{\mr{as}})_{99}=\cos^{2}\varphi+\sin^{2}\varphi\cos^{2}\theta\label{aspure}
\end{split}
\end{equation}
and
\begin{equation}
N(\rho_{\mr{as}})=\sin\varphi\cos\varphi\sin\theta \label{negaspure}
\end{equation}
Comparing (\ref{negpure}) with (\ref{negaspure}) we see that
$$
N(\rho_{\mr{as}})\leq N(\Psi_{\theta,\varphi})
$$
Consider now some special cases: \\
\textbf{1.} Let $\theta=\pi/4$ then
$$
\widetilde{\Psi}_{\varphi}=\frac{\sin\varphi}{\sqrt{2}}\,(\e{1}\otimes\e{1}
+\e{2}\otimes\e{2})+\cos\varphi \e{3}\otimes\e{3}
$$
where $0<\varphi<\pi/2$. The states $\widetilde{\Psi}_{\varphi}$
have negativity
$$
N(\widetilde{\Psi}_{\varphi})=\frac{1}{2}\sin\varphi (2\sqrt{2}\cos
\varphi +\sin\varphi)
$$
and
$$
N(\rho_{\mr{as}})=\frac{\sin\varphi \,\cos\varphi}{\sqrt{2}}
$$
which is always greater then zero.\\[2mm]
\textbf{2.} Let $\theta=\pm\pi/2,\, \varphi\in (0,\pi/2)$, then the
states
$$
\Psi_{\varphi}^{\pm}=\pm\sin\varphi\,\e{2}\otimes\e{2}+\cos\varphi\,
\e{3}\otimes\e{3}
$$
are stable during the evolution, and
$$
N(\rho_{\mr{as}})=N(\Psi_{\varphi}^{\pm})=\cos\varphi\,\sin\varphi
$$
\textbf{3.} Let $\theta=0$, then the states
$$
\Psi_{\varphi}^{0}=\sin\varphi\,\e{1}\otimes\e{1}+\cos\varphi\,
\e{3}\otimes\e{3},\quad \varphi\in (0,\pi/2)
$$
are entangled with negativity
$N(\Psi_{\varphi}^{0})=\sin\varphi\,\cos\varphi$, but
$\rho_{\mr{as}}=\e{3}\otimes\e{3}$ is separable.\\[2mm]
 When the initial
state is the isotropic $W$, then the nonvanishing matrix elements of
$W_{\mr{as}}$ are
\begin{equation}
\begin{split}
&(W_{\mr{as}})_{55}=\frac{1}{9}\,(1+2p),\quad
(W_{\mr{as}})_{59}+\frac{p}{3}\\
&(W_{\mr{as}})_{66}=\frac{2}{9}\,(1-p),\quad
(W_{\mr{as}})_{88}=\frac{2}{9}\,(1-p)\\
&(W_{\mr{as}})_{99}=\frac{2(2+p)}{9}
\end{split}
\end{equation}
and
\begin{equation}
N(W_{\mr{as}})=\W{0}{p\leq
\frac{2}{5}}{\frac{1}{9}\,(5p-2)}{p>\frac{2}{5}}
\end{equation}
So for $p>\frac{2}{5}$, asymptotic states are still entangled.
\section{System of atoms of type (I). Numerical results}
When we consider two atoms of type (I), the analysis of evolution of
the system is much more involved. For distant atoms the master
equation reads
\begin{equation}
\frac{d\rho}{dt}=L_{I}^{AB}\rho\label{meI}
\end{equation}
with
\begin{equation}
\begin{split}
L_{I}^{AB}\rho&=\frac{1}{2}\,\ga{1}(2\sa{31}{A}\rho\sa{13}{A}-
\sa{11}{A}\rho-\rho\sa{11}{A})\\
&+\frac{1}{2}\,\ga{2}(2\sa{32}{A}\rho\sa{23}{A}-\sa{22}{A}\rho-\rho\sa{22}{A})\\
&+\frac{1}{2}\,\ga{12}(2\sa{31}{A}\rho\sa{23}{A}-\sa{21}{A}\rho-\rho\sa{21}{A})\\
&+
\frac{1}{2}\,\ga{12}(2\sa{32}{A}\ro\sa{13}{A}-\sa{12}{A}\rho-\rho\sa{12}{A})\\
&+\frac{1}{2}\,\ga{1}(2\sa{31}{B}\rho\sa{13}{B}-\sa{11}{B}\rho-\rho\sa{11}{B})\\
&+\frac{1}{2}\,\ga{2}(2\sa{32}{B}\rho\sa{23}{B}-\sa{22}{B}\rho-\rho\sa{22}{B})\\
&+\frac{1}{2}\,\ga{12}(2\sa{31}{B}\rho\sa{23}{B}-\sa{21}{B}\rho-\rho\sa{21}{B})\\
&+
\frac{1}{2}\,\ga{12}(2\sa{32}{B}\ro\sa{13}{B}-\sa{12}{B}\rho-\rho\sa{12}{B})
\end{split}\label{gen1AB}
\end{equation}
where $ \sa{kl}{A},\, \sa{kl}{B}$ for $ k,l=1,2,3$ are defined as in
Sect. IV. Notice that in the present section, the vectors $\e{1},\,
\e{2}$ and $\e{3}$ represent states $\ket{1},\, \ket{2}$ and
$\ket{3}$ respectively and cross damping rate $\ga{12}$ is given by
(\ref{g12}).
\begin{figure}[b]
\centering
{\includegraphics[height=70mm,angle=270]{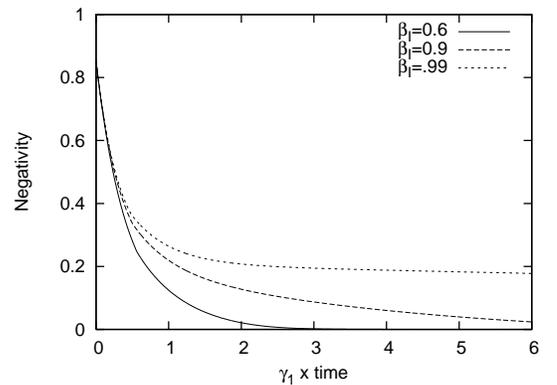}}\caption{Time-dependence
of negativity for pure initial state (\ref{purestates}) with
$\varphi=\pi/6,\, \theta=\pi/8$, for $\ga{2}/\ga{1}=0.9$ and
different values of $\beta_{I}$}
\end{figure}
\begin{figure}[t]
\centering
{\includegraphics[height=70mm,angle=270]{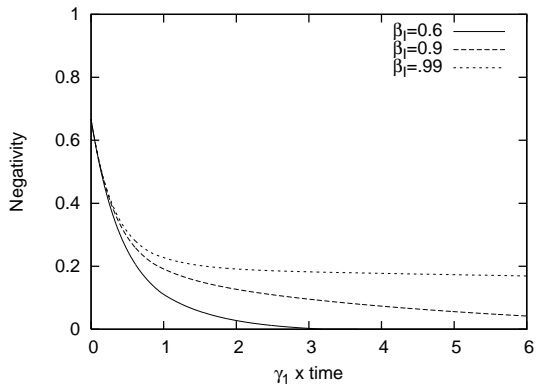}}\caption{Time-dependence
of negativity for isotropic initial state (\ref{W}) with $p=3/4$,
for $\ga{2}/\ga{1}=0.9$ and different values of $\beta_{I}$}
\end{figure}
\par
From (\ref{meI}) we obtain a system of differential equations for
matrix elements of density matrix $\rho$ which we solve numerically.
The results can be used to analyse time evolution of negativity of
given initial state. Since in a single three-level atom the effect
of quantum interference can be observed in both systems (I) and (II)
\cite{ficek}, we expect that also in this model the presence of
interference will slow down the process of disentanglement.

\begin{figure}[t]
\centering
{\includegraphics[height=70mm,angle=270]{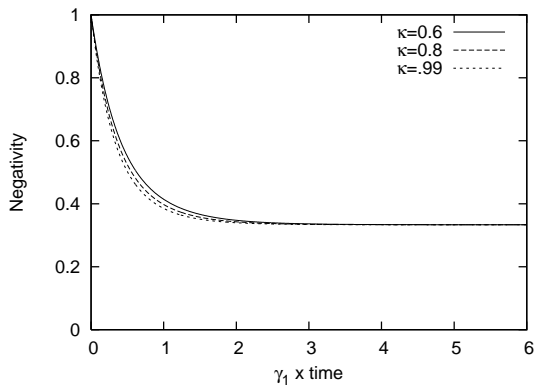}}\caption{Time-dependence
of negativity for initial state (\ref{max}) in the case of maximal
interference and for different values of $\kappa=\ga{2}/\ga{1}$}
\end{figure}
Numerical results entirely confirm these expectations. In
particular, we show that when $\beta_{I}<1$, all  initial states
evolve to the asymptotic state $\ket{\e{3}}\bra{\e{3}}$, so
disentangle asymptotically, but as in the previous case, the rate of
disentanglement depends on quantum interference: for larger
$\beta_{I}$ the process of disentanglement is slower.  FIGS. 6 and 7
represent the results of calculations for pure initial state
(\ref{purestates}) with $\varphi=\pi/6,\, \theta=\pi/8$ and
isotropic state (\ref{W}) with $p=3/4$.
\par
In the case of maximal interference ($\beta_{I}=1$), we also observe
that the system has nontrivial asymptotic states with non-zero
negativity. For example, FIG. 8 represents the results of numerical
calculations of evolution of negativity for maximally entangled
initial state (\ref{max}).
\section{Conclusions}
We have studied dynamical aspects of entanglement in a system ot two
independent three-level atoms in $V$-configuration, interacting with
the vacuum. Spontaneous emission causes decoherence and degradation
of initial entanglement, but quantum interference between principal
transition in each atom can slow down the process of
disentanglement. We have shown this effect using two models of
dynamics, which at the level of single three-level atom predict the
phenomenon of interference. Our results indicate that these two
kinds of quantum evolutions similarly describe dynamical behaviour
of entanglement in the system.

\end{document}